\documentclass[pdftex,twocolumn,epjc3]{svjour3}          % twocolumn
\RequirePackage[T1]{fontenc}
\smartqed  % flush right qed marks, e.g. at end of proof
\RequirePackage{graphicx}
\RequirePackage{mathptmx}      % use Times fonts if available on your TeX system
\RequirePackage{flushend}
\RequirePackage[numbers,sort&compress]{natbib}
\RequirePackage[colorlinks,citecolor=blue,urlcolor=blue,linkcolor=blue]{hyperref}

\journalname{Eur. Phys. J. C}
\usepackage{graphicx}
\usepackage{amssymb}
\usepackage{amsmath}
\usepackage{color}
\usepackage{hyperref}
 \usepackage{tabu}

\def\barray{\begin{array}}
\def\earray{\end{array}}
\def\be{\begin{equation}}
\def\ee{\end{equation}}
\def\ben{\begin{equation} \nonumber}
\def\een{\end{equation}}
\def\ban{\begin{eqnarray*}}
\def\ean{\end{eqnarray*}}
\def\ba{\begin{eqnarray}}
\def\ea{\end{eqnarray}}

\def\({\left(}
\def\){\right)}

\graphicspath{{./fig/}}

\begin{document}
\title{Non-minimal Higgs inflation in the context of warm scenario in the light of Planck data}
\author{Vahid  Kamali\thanksref{e,addr}}
%,Elahe Navaee Nik \thanksref{e1,addr}}

\thankstext{e}{e-mail: vkamali@basu.ac.ir}
%\thankstext{e1}{e-mail: e.navaeenik@gmail.com}
\institute{Department of Physics, Bu-Ali Sina University, Hamedan
65178, 016016, Iran\label{addr}
}
\date{Received: date / Accepted: date}
\maketitle

\begin{abstract}
We investigate the non-minimally Higgs inflaton (HI) model in the context of warm inflation scenario.  Warm little inflaton (WLI) model considers the little Higgs boson as inflaton. The concerns of warm inflation model can be eliminated in WLI model. There is a special case of dissipation parameter  in WLI model $\Gamma=\Gamma_0 T$. Using this parameter, we study the potential of  HI in Einstein frame . Finally we will constrain the parameters of our model using current Planck observational data. 
\end{abstract}


	\maketitle
	%\flushbottom
\section{ Set up and Motivation:}
	Presented by current observational data \cite{Akrami:2018odb,Ade:2015lrj}, our universe is almost homogeneous, isotropic and flat. Inflation explains these concerns in the context of high-energy fundamental physics. There is a quasi-exponential expansion of the universe usually due to a scalar field which slowly rolling down an approximately flat potential \cite{Guth:1980zm, Albrecht:1982wi}. Cosmic microwave background (CMB) data Constrain the properties of inflaton (quanta of inflation) and also present an upper bound for Hubble parameter $H=\frac{\dot{a}}{a}$ ($a(t)$ is scale factor of expanding universe) in inflation epoch $H\leq 10^{-5}M_p\simeq 10^{13}-10^{14} GeV$. This bound of Hubble parameter motivates us to study extension of standard model (SM) of particles, for example: Grand unified theories (GUTs), supersymmetry, extra dimension and string theory. In GUTs there are scalar fields which are required for gauge symmetry breaking e.g. Higgs fields. The potential of these fields often are not flat enough for slow-roll epoch. But the standard model of Higgs boson can be considered as inflaton where the Higgs scalar field non-minimally coupled to gravity \cite{Bezrukov:2007ep} with action \cite{Salopek:1988qh,Kaiser:1994vs,Komatsu:1999mt,Fakir:1990iu}
	\begin{align}
	S_J=\int D^4x\sqrt{-g}\lbrace-\frac{M^2+\xi h^2}{2}R\\
\nonumber	
	+\frac{\partial_{\mu}h\partial^{\mu}h}{2}-\frac{\lambda}{4}(h^2-v^2)^2\rbrace
\end{align}	 
where $\xi$ is non-minimally coupling strength which is related to the 
Brans-Dicke parameter $\omega$ as $\xi=(A\omega)^{-1}$ and vacuum expectation value of the potential $v$ is related to Planck mass $M_p$ as $v^2=\frac{M_p^2}{\xi^2}$. It is possible to transfer the Jordan frame to Einstein frame  by conformal transformation
\begin{align}
	\hat{g}_{\mu\nu}=\Omega^2 g_{\mu\nu},~~~~~~~~~~\Omega=1+\frac{\xi h^2}{M_p^2}
\end{align}
 which leads to non-minimal kinetic term  and  effective potential:
 \begin{align}  \label{po1}
	U(\phi)=\frac{\lambda M_p^4}{4\xi^2}(1+\exp(-\frac{2\phi}{\sqrt{6}M_p}))^{-2}~\\
	\nonumber
	h\simeq\frac{M_p}{\sqrt{\xi}}\exp(\frac{\phi}{\sqrt{6}M_p})~~~~~~~~~~~~~~~~ 
\end{align}    	
   
Recently, warm little inflaton (WLI) model is presented \cite{Bastero-Gil:2016qru}, which shows that pseudo Nambu-Goldenstone boson of a broken gauge symmetry, same as little Higgs scenario for	electroweak symmetry breaking, can be used in the context of warm inflation where the early expansion of the universe may naturally  occur at temperature $T$. The main condition of warm inflation $T>H$ \cite{Berera:1995ie,Berera:1996fm,Hall:2003zp,Moss:1985wn,Berera:1999ws} is sustained by dissipative effects
\begin{align} \label{conser}
	\dot{\rho}_R+4H\rho_R=\Gamma\dot{\phi}^2\\
	\nonumber
	\dot{\rho}_{\phi}+3H(\rho_{\phi}+P_{\phi})=-\Gamma\dot{\phi}^2
\end{align}
	where the canonical form of energy density and pressure of inflaton are
\begin{align}	\label{energy}
	\rho_{\phi}=\frac{1}{2}\dot{\phi}^2+V(\phi),~~~~~~~~P_{\phi}=\frac{1}{2}\dot{\phi}^2-V(\phi).
\end{align}
In WLI model the potential of the inflaton or Higgs boson is protected against thermal correction and quadratic divergences \cite{Bastero-Gil:2016qru}. In this letter we will study the Higgs potential (\ref{po1}) in the context of WLI model by using the dissipation parameter 
 	\begin{align} \label{dissipation}
	\Gamma=\Gamma_0 T
\end{align}
which was presented for WLI model in Ref.\cite{Bastero-Gil:2016qru}.

	\section{Important Theoretical Parameters of Warm Higgs Inflation:}
	In this section we consider a FLRW universe  with canonical energy density in inflation era (\ref{energy}), the basic cosmological equations in the context of the WI \cite{Berera:1995ie, Berera:1996fm} are presented by Friedmann equation 
\begin{equation}\label{2.3}
H^2=\frac{1}{3m_p^2}(\rho_{\phi}+\rho_{\gamma})
\end{equation}
and eq.(\ref{conser}),
%\begin{eqnarray}\label{denn}
%{\dot \rho}_{\phi}+3H(\rho_{\phi}+p_{\phi})=-\Gamma\dot{\phi}^2\\
%\nonumber
%\dot{\rho}_{\gamma}+4H\rho_{\gamma}=\Gamma\dot{\phi}^2~~~~~~~~~
%\end{eqnarray}
where $H=\frac{\dot{a}}{a}$ and $a$ are Hubble parameter and scale factor respectively, dot means derivative with respect to the cosmic time. The model is considered in natural units $\frac{h}{2\pi}=c=1$. Using Eqs.(\ref{conser},\ref{2.3}), we can derive the WI background evolution equation coupled by scale factor in slow-roll limit, $\rho_{\phi}>\rho_{\gamma}, V(\phi)\gg \dot{\phi}^2$.  
\begin{eqnarray}\label{E.O.M}
H^2=\frac{1}{3m_p^2}V  \\
\nonumber
3H(1+Q)\dot{\phi}+\frac{d V}{d\phi}=0\\
\nonumber
\rho_{\gamma}=\frac{\Gamma}{4H}\dot{\phi}^2=C_{\gamma}T^4
\end{eqnarray} 
where $Q=\frac{\Gamma}{3H}$, $T$ is temperature of thermal bath and  $C_{\gamma}=\frac{\pi^2 g_{*}}{30}$ which we presented the degrees of freedom of created massless modes with factor $g_{*}$  .
Now we consider Higgs inflation model in which its potential behaves like (\ref{po1})  
%\begin{eqnarray}\label{shaft}
%V(\phi)=M^4\phi^{2n-2}(\phi^n+m^n)^{\frac{2}{n}-2}
%\end{eqnarray}
%where n is real parameter, $M$ and $m$ are mass-scales \cite{Dimopoulos:2014boa, 
%Dimopoulos:2015aca}.
Slow-roll conditions in the warm inflation scenario are modified, $\epsilon,\vert\eta\vert,\beta<1+Q$ 
\cite{Berera:2008ar,BasteroGil:2009ec}.
The slow-roll parameters are presented by 
\begin{eqnarray}
\epsilon=\frac{M_p}{2•}(\frac{V'}{V})^2\\
\nonumber
\eta=M_p^2(\frac{V''}{V})\\
\nonumber
\beta=M_p^2(\frac{\Gamma'}{\Gamma•}\frac{V'}{V})
\end{eqnarray}
where the prime denotes derivative with respect to scalar field.  
Evolution of inflaton field (\ref{E.O.M}) in term of slow-roll parameters is presented by \cite{Bastero-Gil:2016qru}:
\begin{eqnarray}\label{Number}
\frac{d\phi}{dN}=-\frac{2\sqrt{\epsilon}}{1+Q}
\end{eqnarray}
which is useful in perturbation segment. In Eq.(\ref{Number}) $dN=Hdt$ is the number of e-folds. 
We assumed a spatially-flat, isotropic and homogeneous FLRW universe in the background level, but CMB and large scale structure (LSS) observational data denote  small deviations from isotropic and homogeneity in our universe. These small deviations  motivate us to use linear perturbation theory in cosmology of early time. In the context of general relativity and gravitation, which are bases of cosmology, inhomogeneity grows with time and produces LSS, so it was very small in the very past time. Therefore first order or linear perturbation theory is useful  for scalar field models of inflation epoch. Considering Einstein's equation at the background level, inflaton field evolution in the FLRW universe connects to the evolution of metric components or actually scale factor of this universe, so the evolution of perturbed inflaton field can be considered in the perturbed FLRW geometry. Most general form of the linear perturbations of spatially-flat FLRW metric components are presented by:
\begin{eqnarray}
ds^2=-(1+2C)dt^2+2a(t)D_{,i}dx^i dt~~~~~~~~~~~~~\\
\nonumber
+a^2(t)[(1-2\psi)\delta_{ij}+2E_{,ij}+2h_{ij}]dx^{i}dx^{j}
\end{eqnarray}     
which includes traceless-transverse tensor perturbations $h_{ij}$ and scalar perturbations $C, D, \psi, E$. 
Important perturbation parameters which may be used for comparison between theory and observational data are power spectrum of the curvature perturbation, spectral index and tensor-to-scalar ratio. The primordial power spectrum of warm inflation at the horizon crossing is modified as \cite{Hall:2003zp,Moss:2008yb,BasteroGil:2009ec,Benetti:2016jhf,Bastero-Gil:2016qru} :
\begin{eqnarray}\label{spec}
\Delta_{{\cal R}}(k/k_*) =  P_0(k/k_*) {\cal F} (k/k_*)\\
\nonumber
P_0(k/k_*) \equiv  \left(\frac{ H_{*}^2}{2 \pi\dot{\phi}_*}\right)^2\\
\nonumber
{\cal F} (k/k_*) \equiv  \left(1+2n_* + \frac{2\sqrt{3}\pi
  Q_*}{\sqrt{3+4\pi Q_*}}{T_*\over H_*}\right) G(Q_*).
\end{eqnarray}
where $"*"$ index, denotes the parameters at the horizon crossing. In Eq.(\ref{spec}), $P_0(k/k_*)$ is the power spectrum of cold model of inflation which is modified by  ${\cal F} (k/k_*)$ in the warm scenario of inflation. The modification function $G(Q_*)$ is due to the coupling between inflaton and radiation fluctuations. This function, for linear form of dissipation (\ref{dissipation}) is presented by \cite{Benetti:2016jhf}: 
\begin{eqnarray}
G(Q_*)\simeq 1+ 0.335 Q_*^{1.364}+ 0.0185Q_*^{2.315}.
\end{eqnarray}
On the other hand, $n_*=(\exp(\frac{H}{T})-1)^{-1}$ in Eq.(\ref{spec}) is Bose-Einstein distribution of inflaton field in a radiation bath.  Another perturbation parameters of warm inflation model are  modified as:
\begin{eqnarray}
r= \frac{16\epsilon_{V}}{(1+Q_{\star})^{2}} {\cal F}^{-1} (k/k_*)\\
\nonumber
n_s -1 = \lim_{k\to k_*}   \frac{d \ln \Delta_{{\cal R}}(k/k_*) }{d
  \ln(k/k_*) }
\end{eqnarray}
The modifications of spectral index and tensor-to-scalar ratio are also due to modification of scalar power spectrum.       
We will study our model in strong dissipative regime, $\Gamma\gg 3H$ which agrees with the swampland conjecture \cite{Obied:2018sgi}, in the context of warm inflation \cite{Motaharfar:2018zyb}"
, and in weak dissipative regime, $\Gamma\ll 3H$, by using special dissipative coefficient $\Gamma=\Gamma_0 T$ which is related to the high temperature supersymmetry case \cite{Moss:2006gt} and warm little inflation \cite{Bastero-Gil:2016qru}. In warm inflation, it is straightforward to see that the end of inflation is determined by the condition $\epsilon=1+Q$. We can find the above perturbation parameters for Higgs potential in the context of warm inflation in the strong dissipative regime $Q\gg1$, Table.(\ref{tab:high}).
In this limit the temperature of radiation bath is in form:
\begin{eqnarray}
T=(\frac{V'^2}{4HC_{\gamma}\Gamma_0})^{\frac{1}{5}}
\end{eqnarray}  
for special form of dissipation parameter of warm little Higgs model (\ref{dissipation}).  
\begin{table}[ht]	
\begin{center}
	\begin{tabular}{ | l | c | c |}
		\hline
		$Q\gg 1$ & Theoretical amount  & Constant parameter \\
		 \hline
		$\Delta_R$ & $\Delta_{0R}\exp(-\frac{0.635}{5\sqrt{6}}\frac{\phi}{M_p})$ & $\frac{\Delta_{0R}}{\Gamma_0^{3.51}}=1.1\times 10^{-2.1}$\\
		 \hline
		$n_s-1$ &  $n_{0s}\exp(-\frac{4}{5\sqrt{6}}\frac{\phi}{M_p})$& $\frac{n_{0s}}{\Gamma_0^{-1.6}}=-7.62\times 10^{-2.6}$ \\
		\hline
		$r$ & $r_0(1-n_s)^{-\frac{0.635}{4}}$ & $\frac{r_0}{\Gamma_0^{-5.54}}=2.2\times 10^{-10.08}$ \\
		\hline
		$n_{run}$ & $-\frac{4}{0.635}(1-n_s)^2-n_{0r}(1-n_s)^{\frac{37}{5}}$& $\frac{n_{r0}}{\Gamma_0^{\frac{37}{5}}}=7$ \\
		\hline
	\end{tabular}
		\caption{Important perturbation parameters in the strong dissipative regime  that can be compared with observational data. In these relations we have fixed some parameters as: $\xi^2=10^{8}\lambda$ and $C_{\gamma}=70$.}
		\label{tab:high}
\end{center}
\end{table}\\
In the weak dissipative regime $Q\ll 1$, we also present the theoretical parameters of the model in Table.(\ref{tab:weak}). In this approximation, using dissipation (\ref{dissipation}), we present the new form of temperature as:
 \begin{eqnarray}
T=(\frac{\Gamma_0 V'^2}{36H^3C_{\gamma}})^{\frac{1}{3}}
\end{eqnarray}  
\begin{table}[ht]	
\begin{center}
	\begin{tabular}{ | l | c | c |}
		\hline
		 $Q\ll 1$& Theoretical amount  & Constant parameter \\
		 \hline
		$\Delta_R$ & $\Delta_{0R}\exp(\frac{4}{3\sqrt{6}}\frac{\phi}{M_p})$ & $P_{0R}\simeq3.1\times 10^{-3}(\frac{\lambda^2\Gamma_0}{C_{\gamma}\xi^4})^{\frac{1}{3}}$\\
		 \hline
		$n_s-1$ &  $n_{0s}\exp(-\frac{-2}{\sqrt{6}}\frac{\phi}{M_p})$& $n_{0s}=-1.8$ \\
		\hline
		$r$ & $r_0(1-n_s)^{\frac{4}{3}}$ & $r_0=4.6\times 10^{-1}(\frac{C_{\gamma}\lambda}{6\Gamma_0 \xi^2})^{\frac{1}{3}}$ \\
		\hline
		$n_{run}$ & $n_{r0}(1-n_s)^2$& $n_{r0}=-\frac{3}{4}$ \\
		\hline
	\end{tabular}
		\caption{Important perturbation parameters in the weak dissipative regime  that can be compared with observational data. In comparison with observation we will fix some parameters as: $\xi^2=10^{8}\lambda$ and $C_{\gamma}=70$.}
		\label{tab:weak}
\end{center}
\end{table}\\

These theoretical results can be compared with the Planck observational data.

		\section{ Comparison with observation:}
\begin{figure}
\begin{center}
\includegraphics[scale=0.47]{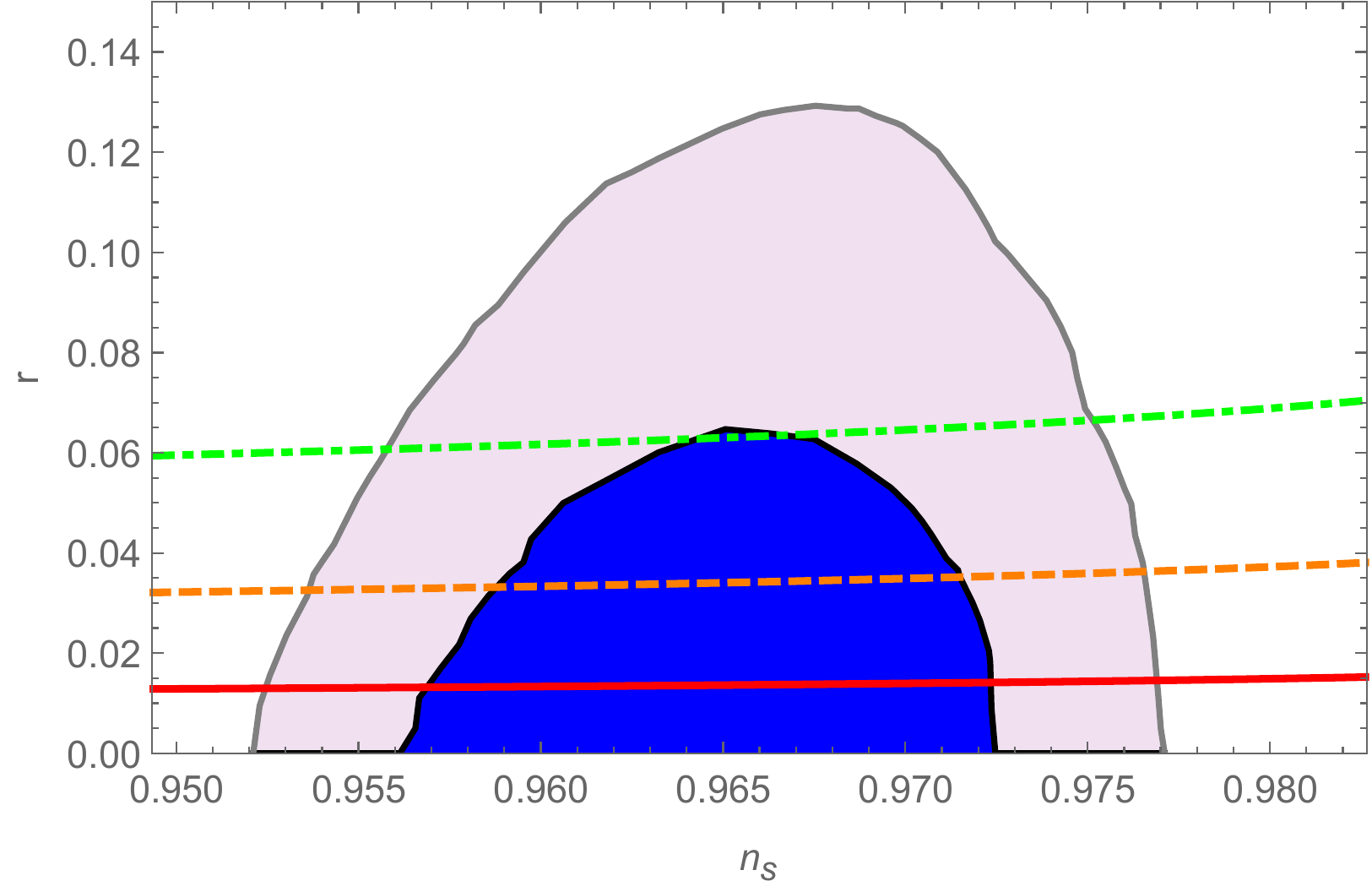}
\end{center}
\caption{$1\sigma$ and $2\sigma$ confidence regions which borrowed from Planck \cite{Akrami:2018odb,Planck:2013jfk},  $r-n_s$ trajectories of the present model in the strong dissipative regime. The solid red, dashed orange and dot-dashed green lines correspond to $\Gamma_0$ values: $1.82\times 10^{-1.64}$, $1.54\times 10^{-1.64}$ and $1.38\times 10^{-1.64}$. In these curves we have fixed $(\xi^2,C_{\gamma})$ as: ($10^8\lambda,70$) and the number of e-folds as: $N_e=60$. }
\label{fig1}
\end{figure}

\begin{figure}
\begin{center}
\includegraphics[scale=0.47]{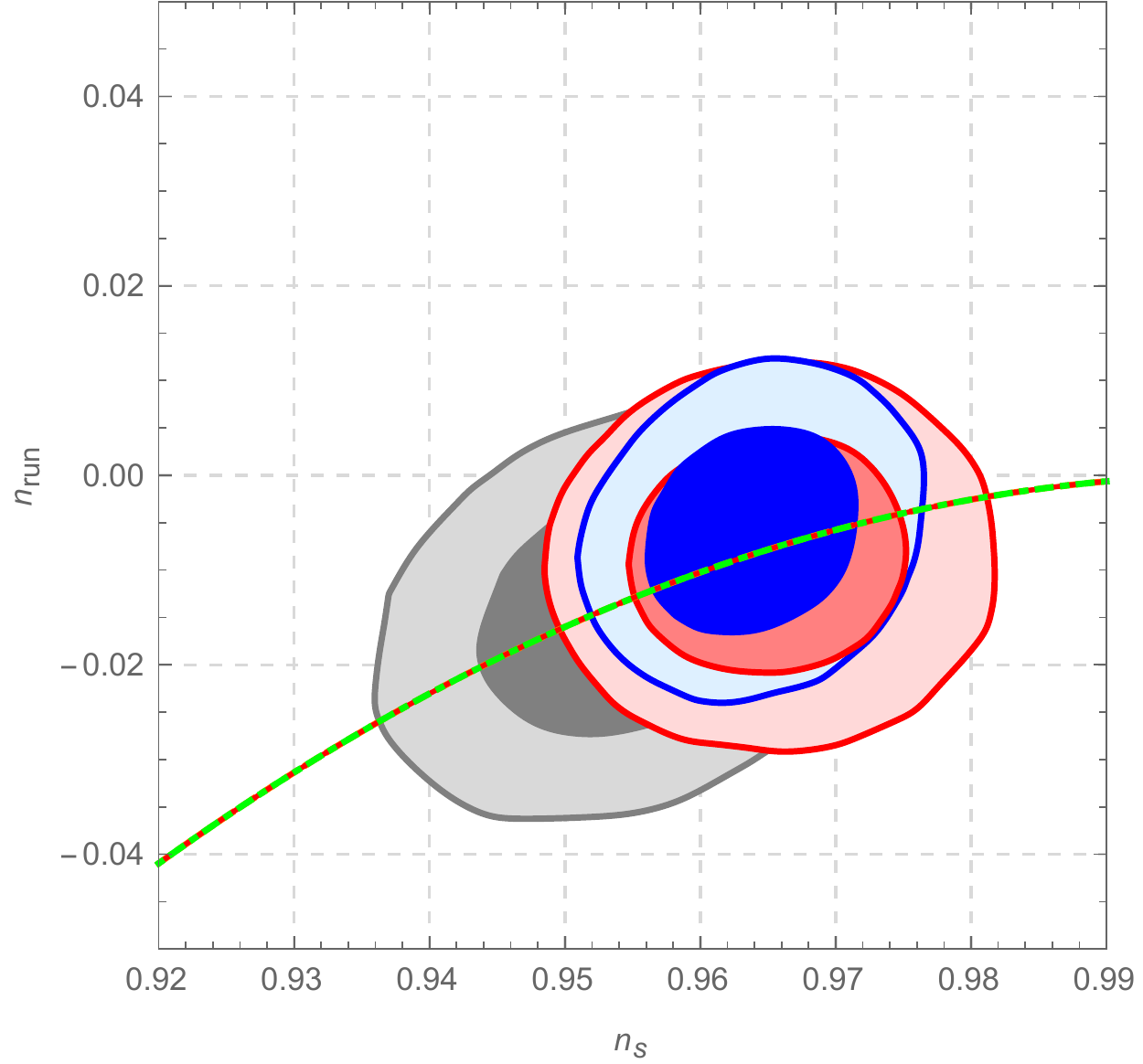}
\end{center}
\caption{$n_s-n_{run}$ diagram. The area corresponds to Planck data and $n_s-n_{run}$ trajectories relate to our model.The solid red, dashed orange and dot-dashed green lines correspond to $\Gamma_0$ values:  $1.82\times 10^{-1.64}$, $1.54\times 10^{-1.64}$ and $1.38\times 10^{-1.64}$. In these curves we have fixed $(\xi^2,C_{\gamma})$ as: ($10^8\lambda,70$) and the number of e-folds as: $N_e=60$. }
\label{fig2}
\end{figure}

%{\bf Comparison with observation:}
Consistency of the perturbation parameters of  our model can be checked by the results of the analysis of \textit{Planck} data sets \cite{Akrami:2018odb,Ade:2015lrj} which indicate the perturbation parameters of  inflation in the slow-roll approximation have limited values:
\begin{eqnarray}\label{lim}
n_s=0.9649\pm 0.0042 ~~~~~~~~~~~~~~~~~~\\
\nonumber
r=\frac{\Delta_T}{\Delta_R}< 0.1,  ~~~~~~~~~~~~~~~~~~~~~~~\\
\nonumber
n_{run}=\frac{dn_s}{d\ln k}=-0.003\pm 0.007~~~~~
\end{eqnarray}   
The small value of running of the spectral index and the upper bound of tensor-to-scalar ratio function $r<0.1$ have been presented by  the results of Planck $95\% CL$ which will be tighter by combined  analysis of  BICEP2/Keck Array BK14/Planck data   as: $r<0.068$ \cite{Akrami:2018odb}. In this section we will try to test the performance of Higgs potential in the context of warm little inflation against the results of observation data (\ref{lim}).
In Fig.(\ref{fig1}) we present the confidence contours in the $(n_s,r)$ plane. Notice that here the tensor-to-scalar ratio in term of spectral index $r(n_s)$ is presented by Table (\ref{tab:high}) in the strong dissipative regime. The value of $\Gamma_0$ is fixed for each trajectory. The curves in this figure are related to  $\Gamma_0$ as: $1.38\times 10^{-1.64}$, $1.58\times 10^{-1.64}$,  and $1.82\times 10^{-64}$ from the top curve to the bottom one. When $\Gamma_0$ decreases the curve is shifted upward. Our model is in $1-\sigma$ confidence level where dissipation parameter has a bound from bottom $\Gamma_0> 1.38\times 10^{-1.64}$ (We have fixed another parameters as:$(\xi^2,C_{\gamma},N_e)=(10^8\lambda,70,60)$).  
 In Fig.(\ref{fig2}), we plotted $n_{run}-n_{s}$ trajectories for some values of $\Gamma_0$  
which have been used in previous figure. There is no difference between these trajectories. The relation between running and spectral index has two parts which the first part is bigger than the second one (see Table.(\ref{tab:high})), so we could not find any difference between $\Gamma_0$ values in this figure. 

\begin{figure}
\begin{center}
\includegraphics[scale=0.47]{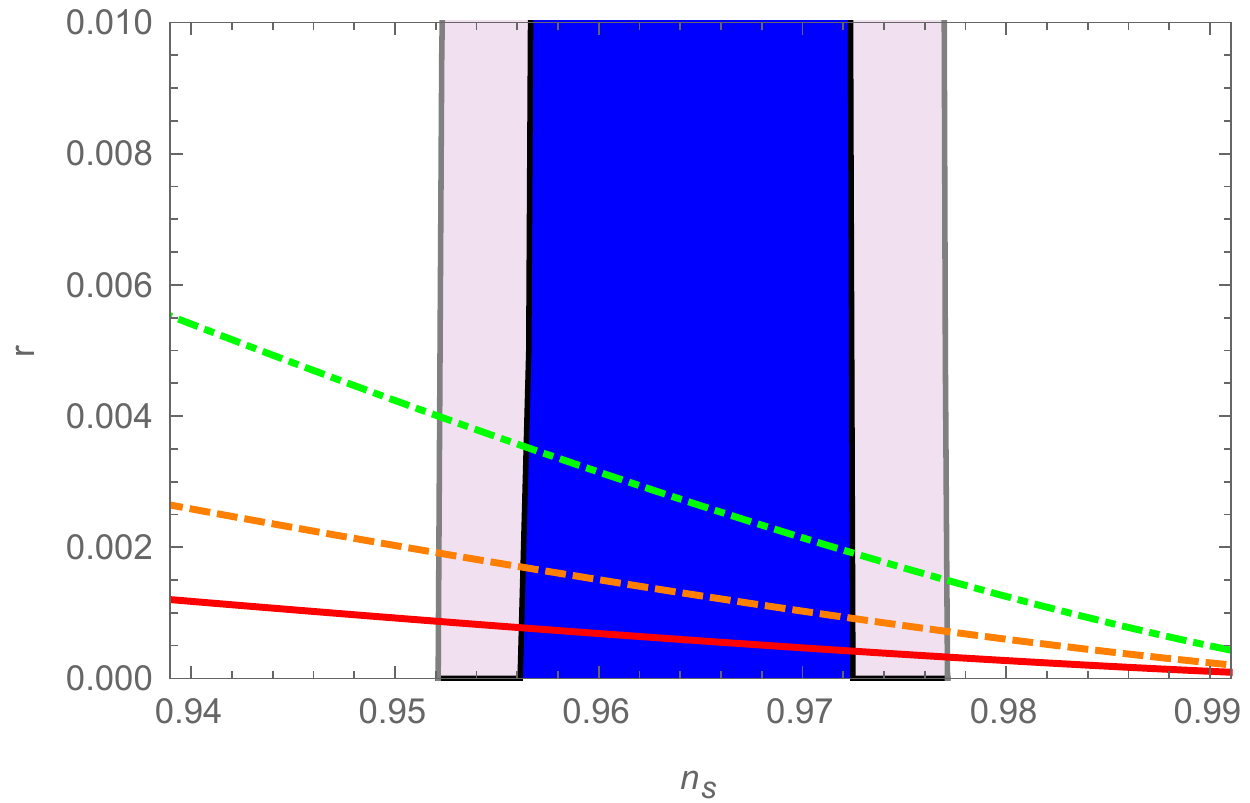}
\end{center}
\caption{ $r-n_{s}$ diagram. The area corresponds to Planck data and $r-n_s$ trajectories relate to our model. The solid red, dashed orange and dot-dashed green lines are  correspond to dissipation parameter $\Gamma_0$ as : $8\times 10^{-7}$, $ 4\times 10^{-7}$, $ 2\times 10^{-7}$. There is a constrain of dissipation parameter $\Gamma_0>2\times 10^{-20}$ for the curves inside the 1-$\sigma$ Planck contour. In these curves we have fixed $(\xi^2,C_{\gamma})$ as: ($10^8\lambda,70$) and the number of e-folds as: $N_e=60$.}
\label{fig3}
\end{figure}

   In Fig.(\ref{fig3}) we present the confidence contours in the $(n_s,r)$ plane. Notice that here the tensor-to-scalar ratio in term of spectral index $r(n_s)$ is presented by Table (\ref{tab:weak}) in the weak dissipative regime. The value of $(\frac{6\Gamma_0\xi^2}{C_{\gamma}\lambda})^{\frac{1}{3}}$ is equal to $\frac{T}{H}=(\frac{6\Gamma_0\xi^2\epsilon}{C_{\gamma}\lambda})^{\frac{1}{3}}$ at the end of inflation in the weak dissipative regime ($\epsilon=1$) which will be fixed for each trajectory. The curves in this figure are related to the value of these coefficients as: $2$, $4$,  and $8$ from the top curve to the bottom one where we have fixed $(\xi^2,C_{\gamma},N_e)$ as: ($10^8\lambda,70,60$). When $\Gamma_0$ increases the curve is shifted downward. The curves are inside of $1-\sigma$ confidence level where $\Gamma_0>2\times 10^{-20}$.
\begin{figure}
\begin{center}
\includegraphics[scale=0.47]{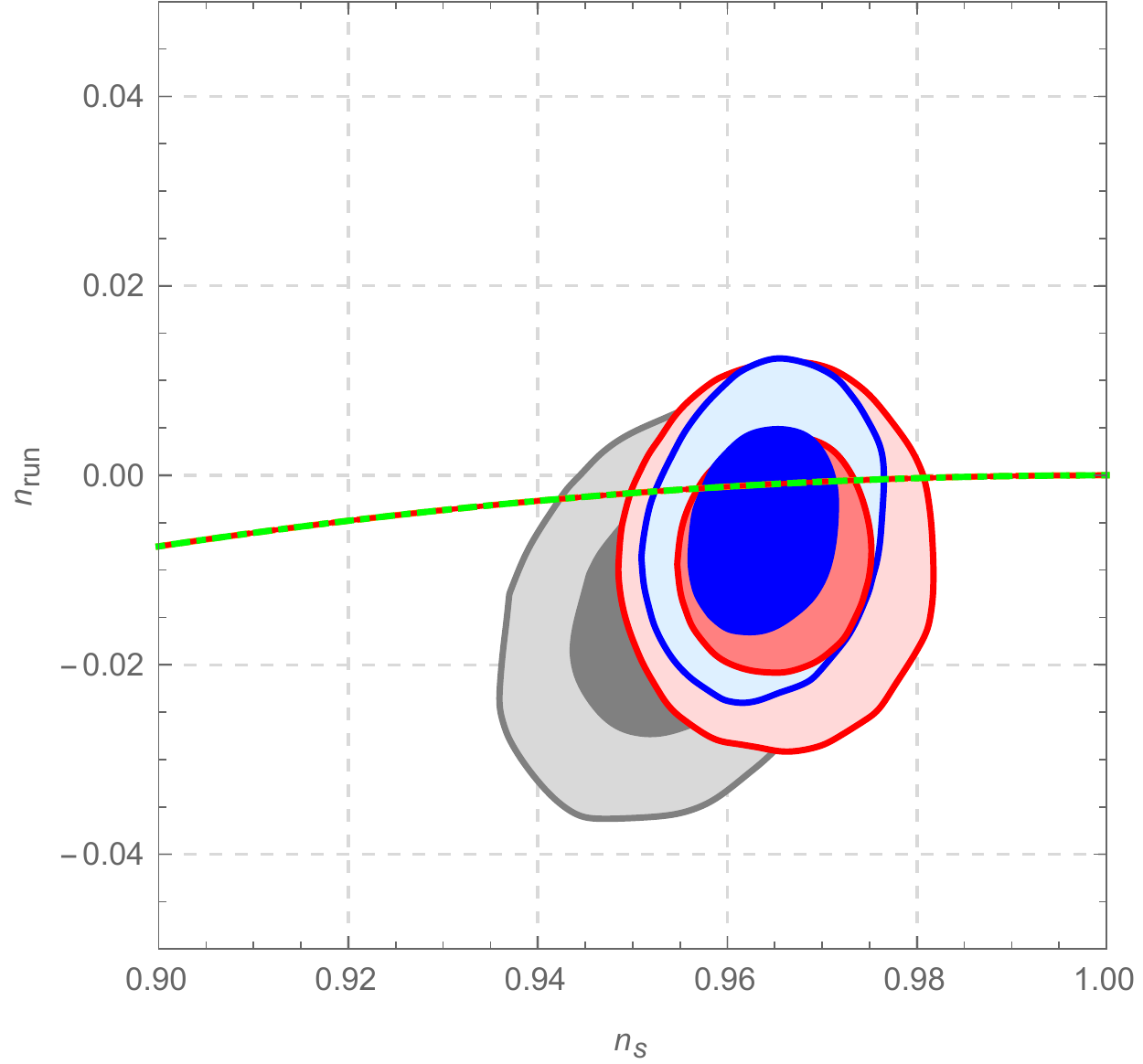}
\end{center}
\caption{ $n_s-n_{run}$ diagram. The area corresponds to Planck data and $n_s-n_{run}$ trajectories relate to our model. The solid red, dashed orange and dot-dashed green lines are  correspond to dissipation parameter $\Gamma_0$ as :  $8\times 10^{-7}$, $ 4\times 10^{-7}$, $ 2\times 10^{-7}$. In these curves we have fixed $(\xi^2,C_{\gamma})$ as: ($10^8\lambda,70$) and the number of e-folds as: $N_e=60$.}
\label{fig4}
\end{figure}
In Fig.(\ref{fig4}), we plotted $n_{run}-n_{s}$ trajectories.   
There is no difference between these trajectories. The relation between running and spectral index is presented by  Table.(\ref{tab:high}) which the coefficient $n_{run0}$ of this relation is constant, so we could not find any difference between trajectories with fixed coefficient  $n_{run0}$ value in figure (\ref{fig4}). 
In Figs.(\ref{fig1}) and (\ref{fig3}) the curves of our model, in the strong and weak dissipative regimes, can be compared with $68\%$ and $95\%$ confidence regions from Planck 2018 results (TT+TE+EE+lowE+lensing+BK14+BAO data)\cite{Akrami:2018odb,Ade:2015lrj} at $k_{*}=0.05 Mpc^{-1}$.  
%\section{Comparison with other models} 
%{\bf Comparison with other models}
Bellow we will compare the current predictions with those 
of viable literature potentials. 
This can help us to understand the variants  
of the warm Higgs-little  inflation model
from the observationally viable inflationary scenarios.
\begin{itemize}
\item The Starobinsky or $R^{2}$ inflation model \cite{staro}:
In Starobinsky inflation model the asymptotic behavior 
of the effective potential is presented as 
$V(\phi)\propto\lbrack1-2\mathrm{e}%
^{-B\phi/M_{pl}}+\mathcal{O}(\mathrm{e}^{-2B\phi/M_{pl}})]$
which provides the following  predictions in the slow-roll limit\cite{Muk81,Ellis13}:
$r\approx 8/B^{2}N^{2}$ and ,$n_{s}\approx 1-2/N$  where $B^{2}=2/3$.
%Furthermore, for Starobinsky inflation following the notations of
%\cite{Ellis13} we find that the running spectral index is given by
%$n_{s}^{\prime}\approx-2/N^{2}$. 
Therefore, if we select $N=50$ then we obtain
$(n_{s},r)\approx(0.96,0.0048)$. For $N=60$ we
find $(n_{s},r)\approx(0.967,0.0033)$. 
It has been found that the Planck data \cite{Akrami:2018odb,Ade:2015lrj}
favors the Starobinsky inflation. Obviously, our results 
(see figures \ref{fig1} and \ref{fig2}) are consistent with those of $R^{2}$ inflation.

\item The Standard Higgs boson as the inflaton\cite{Bezrukov:2007ep}:
In Higgs inflation model the  behavior 
of the effective potential is exponentially flat 
$U(\phi)=\frac{\lambda M_p^4}{4\xi^2}(1+\exp(-\frac{2\phi}{\sqrt{6}M_p}))^{-2}$
where the $1\ll\xi\ll\ll 10^{17}$. This form of potential provides the following  perturbation parameters  in the slow-roll limit\cite{Bezrukov:2007ep}:
$r\approx 192/(4N+3)^2$ and ,$n_{s}\approx 1-8(4N+9)/(4N+3)^2$ 
%Furthermore, for Starobinsky inflation following the notations of
%\cite{Ellis13} we find that the running spectral index is given by
%$n_{s}^{\prime}\approx-2/N^{2}$. 
Therefore, if we select  $N=60$ we
find $(n_{s},r)\approx(0.97,0.0033)$. 
It has been found that the Planck data \cite{Akrami:2018odb,Ade:2015lrj}
favors the non-minimal Higgs inflation. Obviously, our results 
(see figures \ref{fig1} and \ref{fig3}) are consistent with those of non-minimal Higgs inflation.
\item The chaotic  inflation \cite{Linde}:
In this important model of inflation the form of the  potential is presented by
$V(\phi) \propto \phi^{k}$. 
The  slow-roll
parameters for this model are presented as $\epsilon=k/4N$, $\eta=(k-1)/2N$
which leads to main perturbation parameters  $n_{s}=1-(k+2)/2N$ and $r=4k/N$.
Using special  case $k=2$ and $N=50$, we present  
$n_{s}\simeq 0.96$ and $r\simeq 0.16$. For 
$N=60$ we find $n_{s}\simeq 0.967$ and $r\simeq 0.133$. 
It has been found that the 
monomial potentials with $k\ge 2$ are not  in agreement with the Planck
data \cite{Akrami:2018odb,Ade:2015lrj}. 
%Using $k=2$ and $N=50$ we present  
%$n_{s}\simeq 0.96$ and $r\simeq 0.16$. For 
%$N=60$ we find $n_{s}\simeq 0.967$ and $r\simeq 0.133$. 
%It is interesting to note that 
%this model also
%corresponds to the results of intermediate inflation \cite{Barrow:1990vx,Barrow:1993zq,Barrow:2006dh,Barrow:2014fsa} with
%Hubble rate during inflation which is given by $H\propto t^{k/(4-k)}$ with
%$n_{s}=1-(k+2)r/8k$ and $k=-2$ case gives $n_{s}=1$ exactly to the first order. 
%and $n_{s}^{^{\prime}}=-2(n_{s}-1)^{2}/(k+2)$. 
\item Hyperbolic model of inflation \cite{Basilakos:2015sza}:
In hyperbolic inflation the potential is
presented by $V(\phi) \propto \mathrm{sinh}^{b}(\phi/f_{1})$.
Initially, this form of the potential was proposed for dark energy at the late time \cite{Rubano:2001xi}.
This  potential of scalar field has been investigated back in the inflationary era \cite{Basilakos:2015sza} .
The slow-roll parameters are presented by
$$
\epsilon=\frac{b^{2}M_{pl}^{2}}{2f_{1}^{2}}\mathrm{coth}^{2}(\phi/f_1), \label{ee1}%
$$
$$
\eta=\frac{bM_{pl}^{2}}{f_{1}^{2}}\left[  (b-1)\mathrm{coth}^{2}(\phi/f_1)+1\right]
$$
and 
$$
\phi=f_1\;\mathrm{cosh}^{-1}\left[  e^{NbM_{pl}^{2}/f^{2}}\mathrm{cosh}%
(\phi_{end}/f_1)\right]  . \label{efold2}%
$$
where $\phi_{end}\simeq\frac{f}{2}\mathrm{ln}\left(  \frac{\theta+1}{\theta
-1}\right)$. Using observational data, it has been  constrained the parameters of this model.  
$n_{s}\simeq 0.968$, $r\simeq 0.075$, $1<b \le 1.5$ 
and $f_1\ge 11.7M_{pl}$ \cite{Basilakos:2015sza}.
%\item Other models of inflation with exponential form of the potential:
%The origin of brane scenario \cite{Dvali:2001fw,GarciaBellido:2001ky} which is motivated by the physics of 
%extra dimensions  and, on the other hand, the exponential \cite{Goncharov:1985yu,Dvali:1998pa} 
%inflationary models are which motivated by the physics of 
%extra dimentions. 
It has been found in our study that warm Higgs inflation model is in agreement
with the Planck data for some amounts of dissipation coefficient $\Gamma_0$
although using observational data the Starobinsky model of inflation 
is the winner in comparison \cite{Akrami:2018odb,Ade:2015lrj}.
\end{itemize}

		\section{ Conclusions:}
		In this work, 
		we studied the observational signatures of warm inflation 
		in the Cosmic Microwave Background data given by Planck2015.
		We utilized the paradigm of warm inflation with a Higgs scalar field 
		which is non-minimally coupled to gravity. Within this framework at first, we provided the slow-roll parameters and the 
		power spectrum of scalar and tensor fluctuations respectively.
		Second, we checked the performance of  warm Higgs inflationary model
		against the data provided by Planck2015 data and we found a class 
		of patterns which are consistent with the observations. Finally we compare our model with current predictions with those 
of viable literature potentials. 
		
		%In this  work we investigated the tachyon inflation on the brane in the context of 
		%a spatially flat Friedmann-Robertson-Walker  universe. We adopted a specific form of scale factor
		%from Barrow \cite{Barrow:1996bd} solutions, namely logamediate scale  factor.
		%Within this context, we estimated analytically the  
		%slow-roll parameters potential of the model and compare  predictions  
		%with those of other famous inflationary models in the literature.
		%Confronting the model against
		%the latest observational data, we found that the tachyon 
		%inflationary model on the brane  
		%is consistent with the results presented in \emph{Planck 2015} within
		%$1\sigma$ uncertainties. 
		
		\section{acknowledgement}
		I want to thank the Physics School of IPM, which I am a Resident Researcher there, and my colleagues from IPM: Mohammad Mehdi Sheik-Jabbar, Nima Khosravi, Amjad Ashoorioon, Ali Akbar Abolhasani and Seyed Mohammad Sadegh Movahed for some valuable discussions. I want also thanks to my colleagues in BASU: Mohammad Malekjani, Ahmad Mehrabi for some discussions on cosmological perturbation.
		%\begin{thebibliography}{999}
		%\bibitem{general dissipative coefficient}Y. Zhang, JCAP {\bf0903}, 023 (2009) \href{https://arxiv.org/abs/0903.0685}{hep-ph/0903.0685}
		%\bibitem{temperature at the end of inflation} J. Mielczarek, Phys. Rev. D {\bf 83}, 023502 (2011), \href{https://arxiv.org/abs/%1009.2359}{astro-ph/1009.2359}
		%\end{thebibliography}

		% \bibliographystyle{apsrev4-1}
		
		%\bibliographystyle{JHEP}
		\bibliographystyle{apsrev4-1}
		\bibliography{ref}
	\end{document}